\documentclass[aip,jvst,reprint]{revtex4-1}
\usepackage{graphicx}
\usepackage{ulem}
\usepackage{amssymb}
\usepackage{amsmath}
\usepackage{wasysym}
\usepackage{relsize}

\begin{document}


\newcommand{\be}{\begin{equation}}
\newcommand{\ee}{\end{equation}}
\newcommand{\bea}{\begin{eqnarray}}
\newcommand{\eea}{\end{eqnarray}}
\newcommand{\Tbar}{{\bar{T}}}
\newcommand{\En}{{\cal E}}
\newcommand{\K}{{\cal K}}
\newcommand{\U}{{\cal U}}
\newcommand{\GC}{{\cal \tt G}}
\newcommand{\Lop}{{\cal L}}
\newcommand{\DB}[1]{\marginpar{\footnotesize DB: #1}}
\newcommand{\q}{\vec{q}}
\newcommand{\kt}{\tilde{k}}
\newcommand{\Lopn}{\tilde{\Lop}}
\newcommand{\noi}{\noindent}
\newcommand{\ovn}{\bar{n}}
\newcommand{\ovx}{\bar{x}}
\newcommand{\ovE}{\bar{E}}
\newcommand{\ovV}{\bar{V}}
\newcommand{\ovU}{\bar{U}}
\newcommand{\ovJ}{\bar{J}}
\newcommand{\calE}{{\cal E}}
\newcommand{\ovphi}{\bar{\phi}}
\newcommand{\zt}{\tilde{z}}
\newcommand{\ttl}{\tilde{\theta}}
\newcommand{\nuv}{\rm v}
\newcommand{\ds}{\Delta s}
\newcommand{\fn}{{\small {\rm  FN}}}
\newcommand{\cc}{{\cal C}}
\newcommand{\cd}{{\cal D}}
\newcommand{\tth}{\tilde{\theta}}
\newcommand{\cb}{{\cal B}}
\newcommand{\cg}{{\cal G}}
\newcommand\norm[1]{\left\lVert#1\right\rVert}

\title{Approximate universality in the tunneling potential for curved field emitters - a line charge model approach}

\author{Rajasree Ramachandran}\email{rajasreer@barc.gov.in}
\author{Debabrata Biswas}

\affiliation{
Bhabha Atomic Research Centre,
Mumbai 400 085, INDIA}
\affiliation{Homi Bhabha National Institute, Mumbai 400 094, INDIA}

\begin{abstract}
  Field emission tips with apex radius of curvature below 100nm are not adequately described by the
  standard theoretical models based on the Fowler-Nordheim and Murphy-Good formalisms.
  This is due to the breakdown of the `constant electric field' assumption within the
  tunneling region leading to substantial errors in current predictions. A uniformly applicable
  curvature-corrected field emission theory requires that the tunneling potential be approximately
  universal irrespective of the emitter shape.  Using the line charge model, it is established
  analytically that smooth generic emitter tips approximately follow this universal trend
  when the anode is far away. This is verified  using COMSOL for various emitter shapes
  including the locally non-parabolic `hemisphere on a cylindrical post'. It is also
  found numerically that the curvature-corrected tunneling potential provides an adequate
  approximation when the anode is in close proximity as well as in the presence of other emitters.
\end{abstract}

\maketitle

\section{Introduction}
\label{sec:intro}

Over the past few decades, there has been increased use of field emitter based cathodes in vacuum
nanoelectronics. Such cathodes comprise of sharp tips mainly in the form of nanowires, nanocones
and nanotubes, grown on a flat substrate. 
Owing to their advantages such as
fast switching, high brightness, and small energy spread, such cathodes
are widely used in microscopy, lithography, microwave
amplifiers and X-ray generators \cite{fursey,egorov2020}.

The concentration of electric field lines near the tip is quantified using the apex
field enhancement factor $\gamma_a$, defined as the ratio of the local electric field at the
emitter apex $E_a$ to the macroscopic electric field $E_0$ far away from the sharp tip. 
Field enhancement enables  electron emission to occur at moderate field strengths of a few MV/m
compared to the high field strengths ($ > 3\text{GV/m}$) required for planar surfaces.

Conventional field emission theory\cite{jensen_book,liang,forbes_nu,jensen2003,FD2007,DF2008}
is based on the works of Fowler-Norheim (FN) \cite{FN,Nordheim}  and
Murphy-Good (MG) \cite{murphy} which use
the free-electron model and quantum mechanical tunneling through a field dependent potential barrier.
Of the two, the MG theory which includes the image charge potential \cite{Nordheim,murphy}
appears to be closer to physical reality\cite{forbes2019a}
and is increasingly being used\cite{popov2020,shreya_CT}. Since curvature effects do not feature
in the tunneling potential in these theories, they are relevant for 
quasi-planar emitters\cite{cutler93}. The current density at any point on the emitter surface
is expressed as\cite{murphy,forbes_nu,FD2007}

\bea
J_{MG} & = & \frac{A_{\small FN}}{\phi} \frac{E_l^2}{{t_{\small F}}^2}
\exp\left(-B_{\small FN} v_{\small F} \phi^{3/2}/E_l \right) \label{eq:MG} \\
J_{FN} & = & J_{MG} ~\text{with}~t_{\small F} = 1, v_{\small F} = 1  \label{eq:FN}
\eea

\noi
where $E_l$ refers to the local electric field on the emitter surface,
$A_\fn~\simeq~1.541434~{\rm \mu A~eV~V}^{-2}$,
$B_\fn~\simeq 6.830890~{\rm eV}^{-3/2}~{\rm V~nm}^{-1}$ are the conventional FN constants,
$v_{\small F}  =  1 - f_0 + (f_0/6) \ln f_0$, 
$t_{\small F}  =  1 + f_0/9 - \frac{1}{18} f_0 \ln f_0$,
$f_0  \simeq  1.439965~E_l/\phi^2$ and $\phi$ is the work-function.
In the above equations, $J_{MG}$ refers to the Murphy-Good current density
while $J_{FN}$ is the Fowler-Nordheim current density. 
Note that at the apex, $E_l = E_a = \gamma_a E_0$, while
the local field at any other point on the surface depends on the nature of the tip.
For locally parabolic tips, the field near the apex can be determined using the generalized
cosine law \cite{db_ultram},
$E_l = E_a \cos\tilde{\theta}$ where $\cos\tilde{\theta} = (z/h)/\sqrt{(z/h)^2 + (\rho/R_a)^2}$,
$(\rho,z)$ being a point on the surface near the apex of an axially symmetric emitter, $h$ is the height of the emitter and $R_a$ the apex radius of curvature.

The Murphy-Good predictions are in fairly good agreement with current densities
obtained using  tunneling-potentials arrived at analytically
(for solvable models such as the hemiellipsoid) or computed numerically (such as using COMSOL),
when the apex radius of curvature of the emitter is large\cite{db_rrama2019}.
In such cases, the potential due to the applied macroscopic field along the normal
is approximately linear close to the tip surface.
At $R_a = 500$nm
for instance, the error in the net emission current for a hemiellipsoid tip
was found\cite{db_rrama2019} to be around $7\%$ at an apex field of $3$V/nm while
at $R_a = 100$nm, it is slightly greater than $30\%$ and increases beyond $80\%$
at $R_a = 50$nm. Thus, for $R_a \leq 100$nm, there is a need to
introduce curvature-corrections to the Murphy-Good
theory. There is also growing experimental evidence that Eq.~(\ref{eq:MG}) may not 
adequately explain experimental results. In a recent analysis\cite{db_rk_2019} of experimental data
for a single-tip gated  pyramidal emitter\cite{Lee} with apex radius of curvature between 5 and 10nm,
it was found that use of Eq.~(\ref{eq:MG}) to fit the data requires
physical dimensions of the emitter beyond those measured using SEM.
Thus, it is essential to address the role of curvature especially when the
apex radius of curvature is below 100nm.

There are two ingredients in the standard planar field emission theories
that may be a subject of scrutiny for curved emitters. The first is the density of states.
As the emitter height increases and the tip radius becomes smaller, the nature and availability of
electron states can change\cite{CX2017,CX2020}. The second is the effect of
curvature on the tunneling potential for a generic field emission tip. 
If the local field near the
tip is considered to be constant within the tunneling regime, the potential is simply $V_{ext}(s) = E_l s$
where $s$ is the normal distance away from the surface.
An implicit assumption here is that even for curved emitters, the
local field remains approximately perpendicular for about 2nm from the surface.
However, even within this assumption, it is possible
that for highly curved emitters, the local
field is no longer constant but drops in magnitude within the extent of the barrier.
In such a scenario, the potential $V_{ext}$ must be nonlinear in $s$ with the
nonlinear terms dependent on the local radius of curvature.
In this paper, we shall investigate whether this non-linear behaviour
of the external potential is approximately universal for generic smooth emitters.

The question of correction to  the potential ($V_{ext}$) due to an applied external electric field
near a sharp tip has been considered earlier\cite{KX,pop_ext}. The first correction 
{\it along the axis} of an axially symmetric emitter was shown to be\cite{KX}

\be
V_{ext}(s) =  E_l s \left[ 1 - \frac{s}{R_a} + \mathcal{O}(s^2) \right]  \label{eq:corr1}
\ee

\noi
where $R_a$ is the radius of curvature at the apex. In Ref. [\onlinecite{pop_ext}], it
was established that for a hemiellipsoid, a hyperboloid and a hemisphere diode, the
first correction (Eq.~\ref{eq:corr1}) holds for other points near the apex as well. 
In addition, a second correction was provided so that for any point near the
apex,

\be
V_{ext}(s) \approx  E_l s \left[ 1 - \frac{s}{R_2}  +  \frac{4}{3}
  \left( \frac{s}{R_2} \right)^2 + \mathcal{O}(s^3) \right]  \label{eq:corr2}
\ee

\noi
where $R_2$ is the second principle radius of curvature at that point on the surface
(at the apex $R_2 = R_a$). In reality, the curvature dependent terms have
coefficients $c_1$ and $c_2$ which are close to unity for points on the surface close
to the apex. Eq.~(\ref{eq:corr2}) was also found to hold for other shapes such as
a rounded cone and cylinder, both of which were modelled numerically using a suitable
nonlinear line charge distribution\cite{pop_ext}.

This paper aims to put Eq.~(\ref{eq:corr2}) on a firm footing by using the nonlinear line
charge model (LCM) to analytically expand the potential using  general considerations
such as the smoothness and local parabolicity of the
emitter tip. In section \ref{sec:LCM}, we shall review the nonlinear LCM and use it
to arrive at the potential variation along the normal at any point on the emitter
surface close to the apex. In section \ref{sec:Num}, we establish numerically that
the potential expansion arrived at in section \ref{sec:LCM} is close to Eq.~(\ref{eq:corr2})
for $s < 2$nm. Our results also show that Eq.~(\ref{eq:corr2}) holds even when the
anode is in close proximity as well as in the presence of other emitters.
A summary and discussions form the concluding section.

\section{The line charge model}
\label{sec:LCM}

The Line Charge Model (LCM) \cite{mesa,pogorelov,JEN_LCM,harris2016,jap16} is a powerful tool to realize various field emitter tip geometries. In recent studies, the LCM has been extensively used to find the field enhancement factor for a single emitter\cite{db_fef} as well as for a finite-sized large area field emitter (LAFE) where the emitters are arranged in an array or placed at random\cite{db_rudra,rr_db_2019}. The presence of anode in close proximity has also been modelled successfully using LCM\cite{db_anode} and it  has been established that anode-proximity effects can counter electrostatic shielding allowing for a tighter packing of emitters\cite{db_rr_2020}. The LCM has also been used to prove the generalized cosine law \cite{physE}  for field variation near the apex of locally parabolic emitter tips  and improving upon the Schottky conjecture\cite{db_schot} for non-vanishing protrusions placed on a macroscopic base.\\

In the LCM, a  field emitter of height $h$ and tip radius $R_a$ mounted on a grounded conducting plane in a parallel-plate diode configuration, is modelled as a vertically aligned line charge and its image in an external electric field $E_0 = V_0/D$ ($V_0$ is the anode potential and $D$ is 
anode-cathode distance). An appropriately chosen line charge density together with the
external electric field produces a zero-potential surface which mimics the grounded curved
emitter on a conducting plane. The line charge density $\Lambda(z)$ can physically be
thought of as the projection of the induced surface charge density $\sigma(z)$,
along the axis of the emitter and can be expressed as,

\be
\Lambda(z)= 2\pi \sigma (z) \rho (z) \sqrt{1+(\frac{d\rho}{dz})^2}.
\ee

\noi
Thus depending on the functional form of $\Lambda(z)$, different geometrical shapes 
can be realized. For example, a linearly varying line charge  $\Lambda(z) = \lambda z$ with
$\lambda$ constant, mimics the shape of a  hemiellipsoidal emitter\cite{pogorelov,jap16}.    

 The potential at any point ($\rho,z$) due to a vertical line charge placed on a
 grounded conducting plane can be expressed as

\be
\begin{split}
V(\rho,z) = & \frac{1}{4\pi\epsilon_0}\Big[ \int_0^L \frac{\Lambda(t)}{\big[\rho^2 + (z - t)^2\big]^{1/2}} dt ~
  - \\
  &  \int_0^L \frac{\Lambda(t)}{\big[\rho^2 + (z + t)^2\big]^{1/2}} dt \Big] + E_0 z \label{eq:pot} \\
  & \;
\end{split} 
\ee

\noi
where $L$ is the extent of the line charge distribution. The parameters defining the line charge distribution including its extent $L$, can, in principle be calculated by imposing the requirement that the potential should vanish on the surface of the emitter.
 In the following, we shall make use of the LCM to explore the nature of the tunneling potential along the normal to the surface of the emitter.
 
 To begin with, the surface of a vertically aligned axially-symmetric emitter $z = z(\rho)$ can be Taylor expanded  about $\rho = 0$. At $\rho = 0$, $z(0) = h$, $dz/d\rho = 0$ and $d^2z/d\rho^2 = R_a^{-1}$, so that close to the apex, the parabolic approximation

\be
z \approx h - \frac{\rho^2}{2 R_a} \label{eq:parabolic}
\ee

\noi
closely follows the surface $z = z(\rho)$. 
For non-parabolic emitters, the second derivative $d^2z/d\rho^2 = 0$ at the emitter tip corresponding
to a flat top with $R_a \rightarrow \infty$. Since our interest here lies in nanotips, the parabolic
approximation holds good near the apex. The extent upto which it holds however depends on
the shapes of endcaps as we shall discuss later.

  The electric field lines are normal to this surface and thus in the direction

\be
\hat{n} = \frac{1}{\sqrt{1 + (\rho_0/R_a)^2}} (\frac{\rho_0}{R_a} \hat{\rho},\hat{z})
\ee

\noi
at  the point ($\rho_0,z_0$) on the surface. The quantities $\hat{z}$ and $\hat{\rho}$
are respectively the unit normal along the cylindrical axis and the
direction perpendicular to it (radial axis).

Our aim is to determine $\vec{E} = (E_{\rho},E_z)$ using Eq.~(\ref{eq:pot})
at an arbitrary point ($\rho,z$) on or outside the surface using

\be
\vec{E} = E_\rho ~\hat{\rho} + E_z ~\hat{z}  =
-\Big( \frac{\partial V}{\partial \rho} \hat{\rho} + \frac{\partial V}{\partial z} \hat{z} \Big).
\ee

\noi
This can be used to determine $\vec{E}.\hat{n}$ (the magnitude
of the electric field in the direction normal to the surface) and subsequently integrated
to determine the potential variation along the normal.

\subsection{Potential expansion for a nonlinear line charge}

A nonlinear line charge density corresponds to a wide class of smooth axially symmetric emitters such as a rounded cone, pyramid or cylindrical post. A general nonlinear line charge density\cite{db_fef,physE} can be expressed as $\Lambda(z) = z f(z)$, where $f(z)$ is in general not a constant. The field components are obtained by differentiating Eq.~(\ref{eq:pot}). Thus,  
\be
\begin{split}
E_\rho = & -\frac{\partial V}{\partial \rho} = - \frac{\rho}{4\pi\epsilon_0} \left[ \int_0^L dt \left\{ \frac{t f(t)}{[\rho^2 + (z+t)^2]^{3/2}} \right. \right. \\
  & \left. \left.  ~~ - \frac{t f(t)}{[\rho^2 + (z-t)^2]^{3/2}} \right\} \right]  \label{eq:Erho_nlin}
\end{split}
\ee

\noi
while

\be
\begin{split}
E_z = & -\frac{\partial V}{\partial z} = - \frac{1}{4\pi\epsilon_0} \left[ \int_0^L dt \left\{ \frac{t (z+t) f(t)}{[\rho^2 + (z+t)^2]^{3/2}} \right. \right. \\
  & \left. \left.  ~~ - \frac{t (z-t) f(t)}{[\rho^2 + (z-t)^2]^{3/2}} \right\} \right]. \label{eq:Ez_nlin}
\end{split}
\ee

As electron emission predominantly occurs from the apex of the emitter, we shall evaluate the field components in the limit $\rho\rightarrow 0$. Thus, by binomial expansion the field components turn out to be
\be
\begin{split}
  E_\rho & =  - \frac{1}{4\pi\epsilon_0} \left[ \int_0^L dt~f(t) \left\{ \frac{t}{(z+t)^3} - \frac{t}{(z-t)^3} \right\}\rho \right.\\
      & \left.  + \int_0^L dt~\frac{3}{2} f(t) \left\{\frac{t}{(z-t)^5} - \frac{t}{(z+t)^5} \right\}\rho^3  + \ldots \right]
\end{split}
\ee

\noi
while

\be
\begin{split}
E_z & =  \frac{1}{4\pi\epsilon_0} \left[ \int_0^L dt~f(t) \left\{ \frac{t}{(z-t)^2} - \frac{t}{(z+t)^2} \right\} \right.\\
  & \left.  + \int_0^L dt~\frac{3}{2} f(t) \left\{\frac{t}{(z+t)^4} - \frac{t}{(z-t)^4} \right\}\rho^2 + \ldots \right].
\end{split}
\ee 

Using partial integration, each of these integrals can be evaluated and keeping terms upto $\rho^2$, the above equations reduces to 

\be
E_\rho \simeq \frac{f(L)}{4\pi\epsilon_0} \frac{2L^3}{(z^2-L^2)^2}(1-\cc_0) \rho \label{eq:Erho_local}
\ee

\be
\begin{split}
E_z \simeq   \frac{f(L) }{4\pi\epsilon_0} & \left[ \frac{2zL}{z^2-L^2}(1-\cc_1)-\ln(\frac{z+L}{z-L})(1-\cc_2) \right. \\
 & \left. - \frac{4zL^3}{(z^2-L^2)^3}(1-\cc_3)\rho^2 \right]  \label{eq:Ez_local}
\end{split}
\ee
where,
\bea
\cc_0 & = & \int_0^L dt~ \frac{f'(t)}{f(L)} \frac{t^3/(z^2 - t^2)^2}{L^3/(z^2 - L^2)^2} \\
\cc_1 & = & \int_0^L dt~ \frac{f'(t)}{f(L)} \frac{t/(z^2 - t^2)}{L/(z^2 - L^2)} \\
\cc_2 & = & \int_0^L dt~ \frac{f'(t)}{f(L)} \frac{\ln((z+t)/(z-t))}{\ln((z+L)/(z-L))} \\
\cc_3 & = & \int_0^L dt~ \frac{f'(t)}{f(L)} \frac{t^3/(z^2 - t^2)^3}{L^3/(z^2 - L^2)^3}.
\eea

\noi
Here $f(L)=\Lambda (L)/L$ and the terms $\cc_0,\cc_1,\cc_2,\cc_3$ arise due to the non linearity in the line charge distribution and are zero otherwise.

Smooth emitters are characterized by a well-behaved charge distribution
and can be expressed as a polynomial function of
degree $n$. Thus, $f(t)$ obeys Bernstein's inequality \cite{bern_ineq}

\be
|f'(x)| \leq \frac{n}{(1 - x^2)^{1/2}} \norm{f}  \label{eq:bern}
\ee

\noi
where $x \in [-1,1]$ and $\norm{f}$ denotes the maximum value of $f$ in this interval. With
$x = t/L$ and applying the inequality, it can be shown that $\cc_k$ are
vanishingly small for sharp emitters for which
$z \simeq h \simeq L$ (see Appendix).
Also, since $\frac{2zL}{z^2-L^2}>\ln(\frac{z+L}{z-L})$ for sharp emitters,
the logarithmic term can be ignored. Hence, the field components turn out to be 

\be
E_\rho \simeq \frac{f(L)}{4\pi\epsilon_0} \frac{2L^3}{(z^2-L^2)^2} \rho \label{eq:Erho_local}
\ee

\be
\begin{split}
E_z \simeq   \frac{f(L) }{4\pi\epsilon_0} & \left[ \frac{2zL}{z^2-L^2} - \frac{4zL^3}{(z^2-L^2)^3}\rho^2 \right].  \label{eq:Ez_local}
\end{split}
\ee

\noi
These field components can be used to evaluate the form of external potential
for a point on the surface of the emitter in the neighbourhood of its apex.

\vskip 0.15 in
$\;$
\subsection{Potential variation along the axis}

It is instructive to first study the potential along the axis of symmetry.
As expected, $E_\rho \rightarrow 0 $ as $\rho \rightarrow 0$. 
With $\rho = 0$ and $z = h + s$ where $s$ is the distance from the apex along the normal,

\bea
E_z & = & \frac{f(L)}{4\pi\epsilon_0} \frac{2zL}{z^2-L^2} \nonumber \\
& = & \frac{f(L)}{4\pi\epsilon_0} \frac{2hL(1+s/h)}{\left[(h^2 - L^2) + 2hs + s^2\right]} \nonumber \\
& \approx & \frac{f(L)}{4\pi\epsilon_0} \frac{2h^2}{(h^2-L^2)} \frac{(1+s/h)}{\left [1+(2hs+s^2)/(h^2-L^2)\right]} \nonumber \\
& = & \frac{f(L)}{4\pi\epsilon_0} \frac{2h/R_a(1+s/h)}{\left[ 1 + 2s/R_a + (s/R_a)(s/h) \right] } \nonumber \\
& \approx & \gamma_a E_0 \left[ 1 + (\frac{1}{h} - \frac{2}{R_a})s + (\frac{4}{R_a^2} - \frac{3}{hR_a})s^2 \right] \nonumber \\
& \approx & E_a \left[ 1  - \frac{2}{R_a} s + \frac{4}{R_a^2} s^2 \right]
\eea

\noi
where we have used $L \approx h$ in the numerator, $\frac{f(L)}{4\pi\epsilon_0} \frac{2h}{R_a} = \gamma_a E_0 = E_a$ which is the electric field at the apex\cite{db_fef}, and $(h^2 - L^2)/h \approx R_a$, the apex radius of curvature\cite{db_fef}. We have also ignored $s/h$ in comparison to unity for a sharp emitter.
Thus the external potential takes the approximate universal form,

\be
V_{ext}(s) = E_a s \left[1 - \frac{s}{R_a} + \frac{4}{3} \left( \frac{s}{R_a}\right)^2  + \mathcal{O}(s^3) \right]
\ee

\noi
along the emitter axis.

\subsection{The potential variation away from  the axis } 

Field emission of electrons predominantly occurs from the close vicinity of the apex. It is thus
necessary to carry out a study on the nature of potential in this region as well. Assuming the field lines to be approximately straight in the tunneling region, let $s$ be the normal distance from a point $(\rho_0,z_0)$ on the emitter surface. Any arbitrary point $(\rho,z)$ along the normal can be expressed as

\bea
\rho & = &\rho_0+ \frac{\rho_0}{R_a}\frac{s}{\sqrt{1+\rho_0^2/R_a^2}} \\
z   & = & z_0+ \frac{s}{\sqrt{1+\rho_0^2/R_a^2}}\\
& \approx &  h - \frac{\rho_0^2}{2R_a}+ \frac{s}{\sqrt{1+\rho_0^2/R_a^2}}.
\eea

\noi
To evaluate the electric field components, the quantity $z^2-L^2$ can be expanded as 
\bea
\begin{split}
z^2 - L^2 & \approx  \left(h - \frac{\rho_0^2}{2R_a}+ \frac{s}{\sqrt{1+\rho_0^2/R_a^2}}\right)^2 -L^2 \\
& =  hR_a \left[ 1 - \frac{\rho_0^2}{R_a^2} + \frac{2 s}{R_a\sqrt{1+\rho_0^2/R_a^2}} + \right.  \\
&~~~ \left. \frac{s^2}{hR_a(1+\rho_0^2/R_a^2)} - \frac{\rho_0^2 s}{h R_a^2 \sqrt{1+\rho_0^2/R_a^2}} +\frac{\rho_0^4}{4hR_a^3} \right]
 \end{split} 
\eea
For high aspect ratio emitters $h\gg R_a, s, \rho_0$ and hence keeping only the dominant terms, the above expression reduces to

\be
z^2 - L^2  \approx  hR_a \left[ 1 - \frac{\rho_0^2}{R_a^2} + \frac{2 s}{R_a} 
 (1-\frac{\rho_0^2}{2R_a^2})\right].
\ee

\noi
Hence the field components after expansion in $\rho_0$ can be expressed as
\bea
E_\rho & \approx & \frac{f(L)}{4\pi\epsilon_0} \frac{2h}{R_a} \left[\frac{\rho_0}{R_a}(1-\frac{3s}{R_a}+\frac{8s^2}{R_a^2})  \right. \nonumber\\
& ~ & ~~~~~~~~~ + \left. \frac{2\rho_0^3}{R_a^3}(1-\frac{9s}{R_a}+\frac{14s^2}{R_a^2}) \right]\\
E_z  & \approx & \frac{f(L)}{4\pi\epsilon_0} \frac{2h}{R_a} \left[(1-\frac{2s}{R_a} + \frac{4s^2}{R_a^2})  \right. \nonumber \\
& ~ & ~~~~~~~~~ - \left. \frac{\rho_0^2}{R_a^2}(1-\frac{5s}{R_a}+\frac{18s^2}{R_a^2}) \right].
\eea 

\noi
In the apex neighbourhood where $\rho_0 \ll R_a$, only the terms upto $\rho_0^2$ will suffice. Thus
the magnitude of the normal electric field is given by
\bea
\begin{split}
|E| = \vec{E}.\hat{n}
  \approx \frac{f(L)}{4\pi\epsilon_0} & \frac{2h}{R_a} \frac{1}{\sqrt{1 + (\rho_0/R_a)^2 }} \left[(1-\frac{2s}{R_a} + \frac{4s^2}{R_a^2}) \right.\\
    &\left.~~~~+ \frac{\rho_0^2}{R_a^2} (\frac{2s}{R_a}-\frac{10s^2}{R_a^2})\right] \\  
 = \frac{E_a}{\sqrt{1 + (\rho_0/R_a)^2 }} 
   & \left[1-\frac{2s}{R_a}(1-\frac{\rho_0^2}{R_a^2})+\frac{4s^2}{R_a^2}(1-\frac{5\rho_0^2}{2R_a^2})\right].
 \end{split}  
\eea
 
\noi
Note that,
\bea
\begin{split}
   \frac{E_a}{\sqrt{1 + (\rho_0/R_a)^2 }} & \approx E_a \frac{(z_0/h)}{\sqrt{(z_0/h)^2 + (\rho_0/R_a)^2}} \\ 
  & =E_a \cos \tilde \theta  = E_l
\end{split}
\eea

\noi
where $\tilde \theta $ corresponds to the launch angle measured from the emitter axis and $E_l$ is the local electric field at the point ($\rho_0,z_0$). Thus,
\be
\begin{split}
|E| = E_l\left[1-\frac{2s}{R_a}(1-\frac{\rho_0^2}{R_a^2})+\frac{4s^2}{R_a^2}(1-\frac{5\rho_0^2}{2R_a^2})\right].
 \end{split} 
\ee

\noi
Hence, finally the external potential is given by
 
 \be
\begin{split} 
V_{ext}(s)  =  E_l s & \left[ 1 - \frac{s}{R_a} (1-\frac{\rho_0^2}{R_a^2})+
 \frac{4~s^2}{3~R_a^2} (1-\frac{5\rho_0^2}{2R_a^2}) \right].
\end{split}
\ee

\noi
Rewriting the above expression in terms of the second principal radius of curvature $R_2 = R_a \sqrt{1+\rho_0^2/R_a^2}$, we finally have
 
 \be
\begin{split} 
V_{ext}(s)  =  E_l s & \left[ 1 - \frac{s}{R_2} (1-\frac{\rho_0^2}{2R_a^2})+
 \frac{4~s^2}{3~R_2^2}  (1-\frac{3\rho_0^2}{2R_a^2}) \right]. \label{eq:LCMeq}
\end{split}
\ee

At a local apex field of 6 V/nm, the current distribution peaks around
$\rho_0 \approx R_a/5$ for an emitter with workfunction $\phi = 4.5$eV.
Thus, the correction terms may be considered small and the
external potential should approximately follow

\be
V_{ext} (s)  \approx   E_l s \left[ 1 - \frac{s}{R_2} + \frac{4}{3} \left(\frac{s}{R_2}\right)^2 \right] \label{eq:final}
\ee

\noi
in the tunneling region. We shall verify this using COMSOL in the following section.

\section{Numerical Results}
\label{sec:Num}

The electrostatic potential can be evaluated numerically for various emitter shapes. In the
following, we shall present our results using the AC/DC module\cite{comsol} of COMSOL v5.4 for a parallel-plate diode geometry
with the emitter mounted on the cathode plate. Due to the axial symmetry of single emitters,
a 2-dimensional domain is used while for emitters on a square lattice,
a 3-dimensional modelling is carried out. 
The emitter is modeled as a perfect electrical conductor which is at ground potential along with the
cathode plane. In the 2-dimensional ($\rho,z$) simulations, the
left boundary is the axis of symmetry while the right boundary
has Neumann boundary condition (zero charge). The top boundary (anode) is Dirichlet with the
potential specified or when the anode is far away, a Neumann boundary condition is used
specifying the surface charge to be $\epsilon_0 E_0$\cite{assis}.
The mesh parameters have been chosen such that the local field at the apex converges
and for solvable models such as the hemiellipsoid with the
anode far away, the values are also compared with analytical results. For 3-dimensional
simulations, similar boundary conditions are used.

We shall interchangeably use the expressions `exact result'
or `numerical result' to refer to the COMSOL result. As with all numerical methods and the finite
nature of the platforms on which they are implemented, the results obtained using COMSOL are
at best a close approximation to what may be considered as exact.

Our aim in the following is to investigate the appropriateness of Eqns.~(\ref{eq:LCMeq})
and (\ref{eq:final}) in modelling field emission for apex radius of curvature in the 5-10nm range
and apex electric fields in the 3-7 V/nm range. It is clear from the nature of the
corrections in Eqns.~(\ref{eq:LCMeq})-(\ref{eq:final}) that curvature-dependent terms in the
external potential diminish in magnitude as the apex of radius of curvature increases.
Thus, the 5nm limit is the ideal testing ground at which it is expected that curvature corrections applied to
current field emission models may work and below which, other phenomenon may come into play.
Similarly, for the tunneling potential, an apex field of 3V/nm is the limit at which
field emission current may be measurable and where the relevance of the cubic term in Eq.~(\ref{eq:final})
may be best investigated. We shall thus stay away from $R_a < 5$nm and $E_a < 3$V/nm.

\subsection{ Anode is far}

We first present the details of the study of external potential for two different geometries (i) Hemiellipsoid on Cylindrical Post (HECP) and (ii) Hemisphere on Cylindrical Post (HCP), both mounted on a cathode plane with anode far away at a distance more than $5$ times the height of the emitter pin. Since these geometries are not analytically solvable, the validity of the formula for external potential is tested using the `exact' results obtained from COMSOL.

\begin{figure}
\includegraphics[scale=.33]{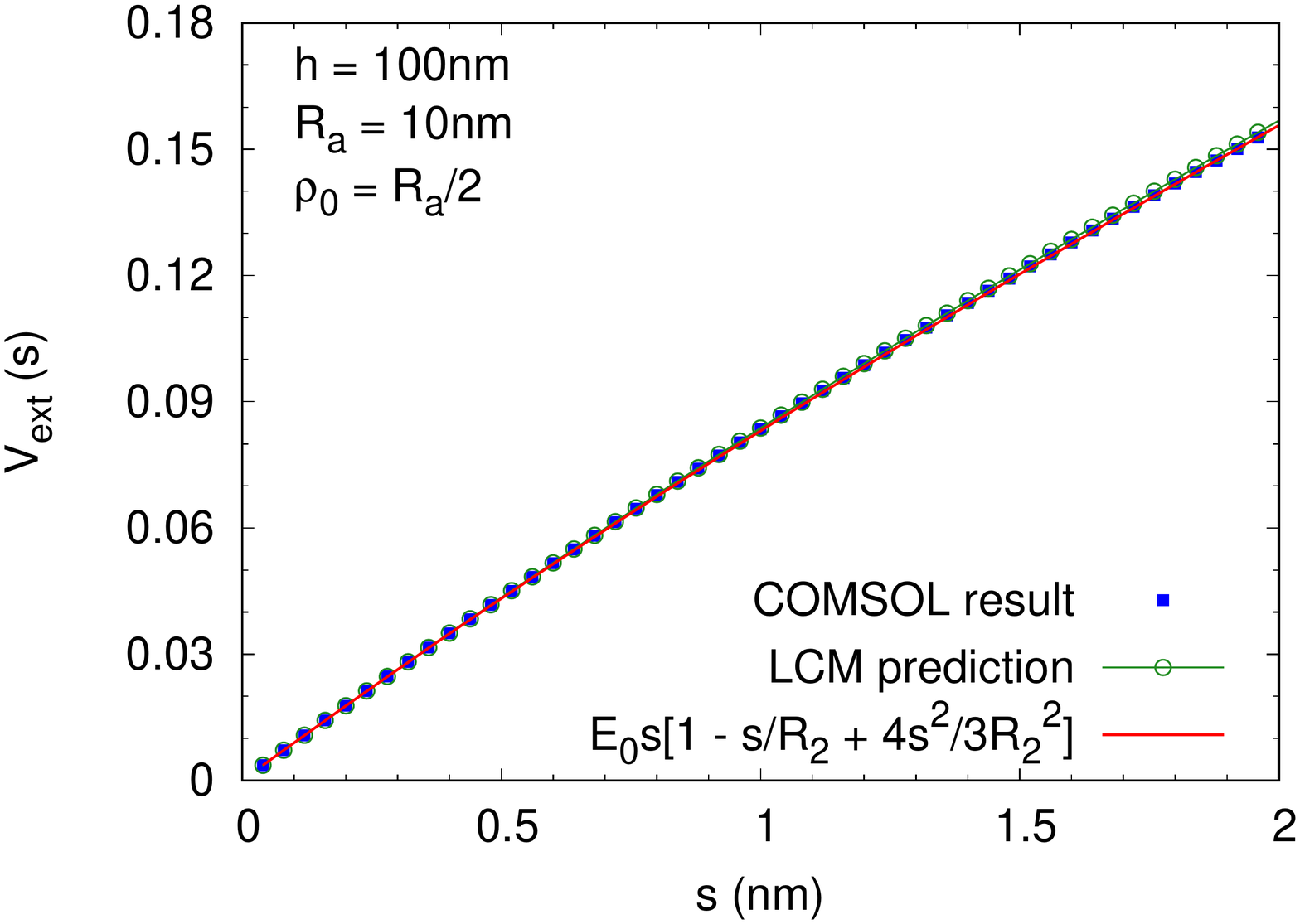} 
\caption{A comparison of external potential obtained using Eq.~(\ref{eq:LCMeq}) (solid squares) and Eq.~(\ref{eq:final}) (open circles) with that of COMSOL (solid curve) for an HECP geometry. The variation is shown
for a point $\rho_0 = R_a/2$ located on a hemiellipsoid of height $h = 100$nm and apex radius $R_a = 10nm$.}
\label{fig:HECP_iso}
\end{figure}

For an HECP emitter, the end cap geometry follows the parabolic approximation of Eq~(\ref{eq:parabolic}) upto $\rho_0 \approx 4R_a/5$. Since field emission at moderate fields ($E_a \in [3,7]$ V/nm) is largely restricted to the region $\rho_0 \leq R_a/2$, we present in Fig.~(\ref{fig:HECP_iso}) a comparison at $\rho_0 = R_a/2$. The total height of the emitter
is $h=100$nm while the height of the cylindrical post is 50nm. The hemiellipsoid endcap has a height 50nm and
apex radius of curvature 10nm.
Clearly, within the tunneling distance (typically $\leq 2 $nm),        
the analytical expression given by Eq.~(\ref{eq:final}) and Eq.~(\ref{eq:LCMeq}) are in good agreement with the exact results.

 \begin{figure}[h]
\includegraphics[scale=.33]{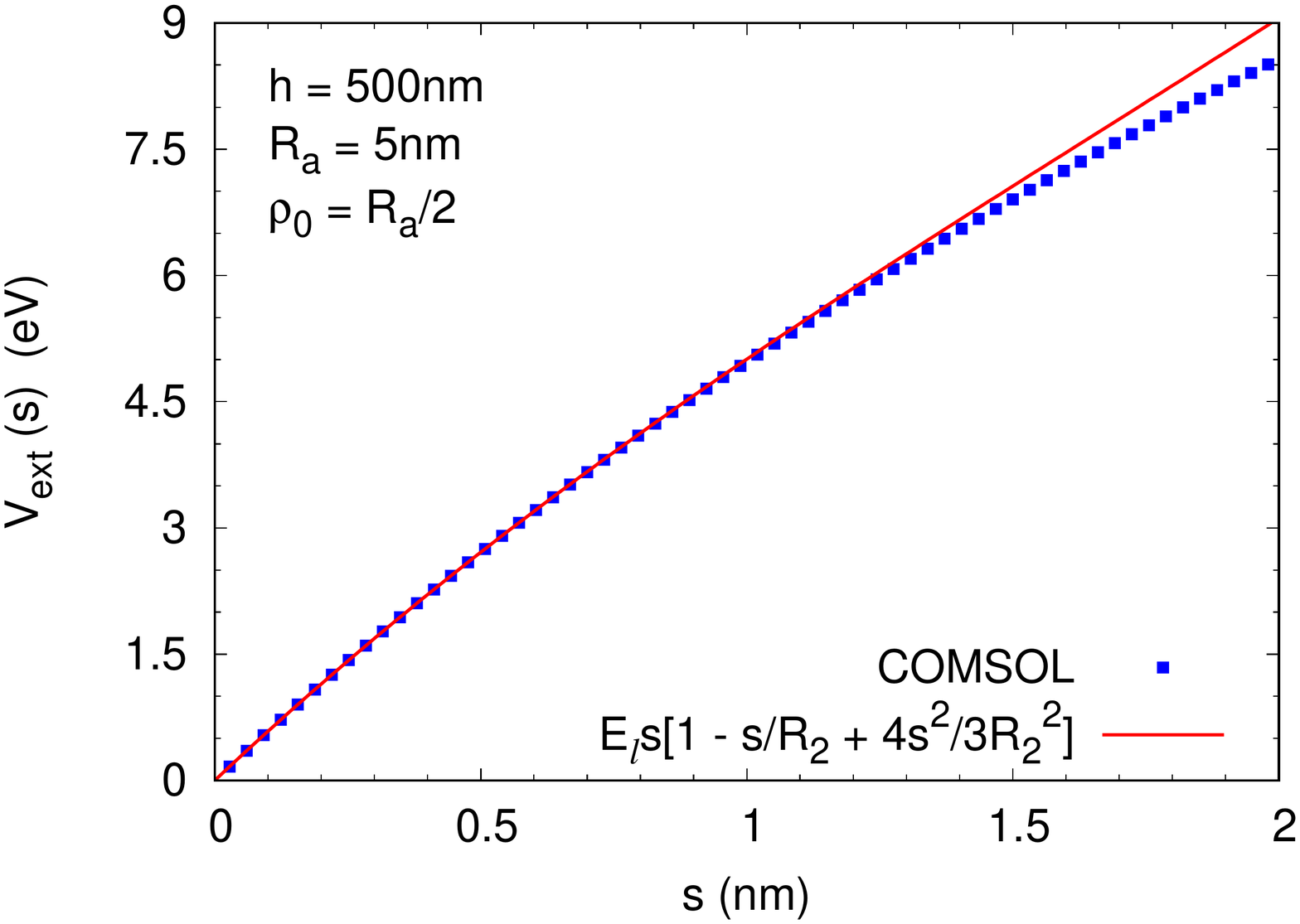}
\caption{The figure shows comparison of  the external potential obtained using Eq.~(\ref{eq:final}) with that of COMSOL for a hemisphere on a cylindrical post (HCP) emitter tip at a point $\rho_0 = R_a/2$.}
\label{fig:HCP_iso}
 \end{figure}

 The hemisphere on a cylindrical post is a special case of the HECP emitter. When the hemiellipsoid endcap height
 is typically five times the apex radius of curvature or greater, the parabolic approximation closely
 follows the emitter shape till $\rho_0 = 4R_a/5$. As the hemiellipsoid height decreases compared to the
 apex radius, this domain of validity shrinks. When the height of the hemiellipsoid equals the
 apex radius (hemisphere endcap), the parabolic approximation follows the hemisphere for $\rho_0 < 2R_a/5$.
 Thus, Eqns.~(\ref{eq:final}) and Eq.~(\ref{eq:LCMeq}) are not expected to as good at $\rho_0 = R_a/2$.
 Fig~(\ref{fig:HCP_iso}) however shows that there is a fair agreement between Eq.~(\ref{eq:final})
 and the exact result even for $R_a = 5$nm and $\rho_0 = R_a/2$. We have also found Eq.~(\ref{eq:final}) to hold
for rounded conical and pyramidal structures as well.

 \subsection{ Anode proximity }

The analytical results presented in section \ref{sec:LCM}
deal with an isolated emitter with the anode far away. While these results  can be analytically extended to
deal with anode-proximity and shielding due to other emitters, we shall instead test the sanctity
of Eq.~(\ref{eq:final}) numerically when the anode is brought near or when the emitter is part of an array.
 
\begin{figure}
\includegraphics[scale=.33]{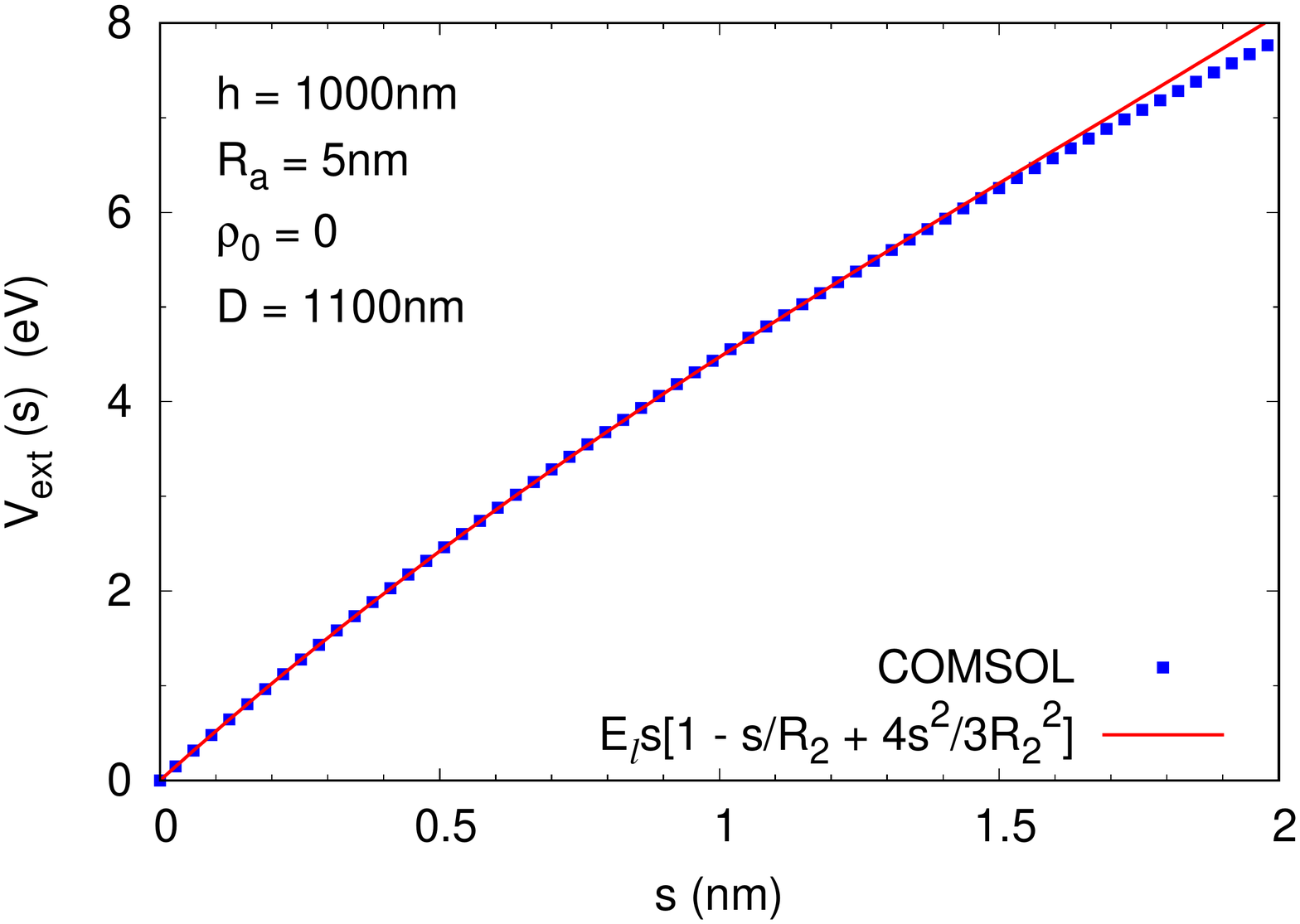}
\caption{A comparison of the external potential  computed using COMSOL (solid squares)
  with Eq.~(\ref{eq:final}) (solid curve)  along the the axis ($\rho_0=0$) of a hemiellipsoidal emitter tip of
  height $h = 1000$nm, tip radius $R_a = 5$nm. The anode is at a distance 100nm from the
  apex.}
\label{fig:ellip_anode_0}
\end{figure}

The presence of anode cannot be neglected in actual field emission current measurements. Also, it is a known fact that the proximity of anode increases  the local electric field at the emitter tip. Hence it is worth investigating how the anode proximity affects the external potential.

\begin{figure}
\includegraphics[scale=.33]{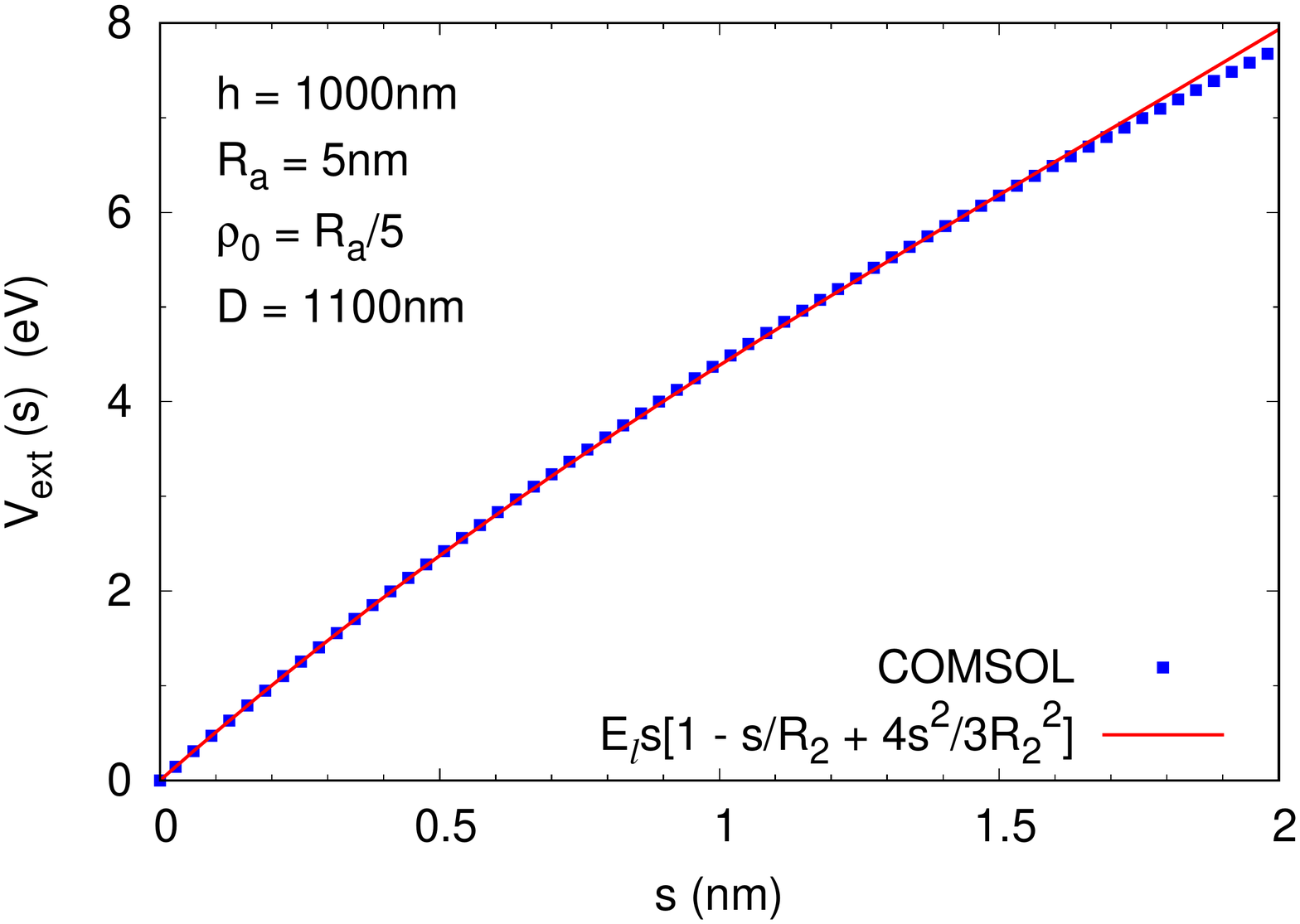}
\caption{As in Fig.~\ref{fig:ellip_anode_0} with $\rho_0 = R_a/5$.}
\label{fig:ellip:anode_Ra_5}
\end{figure}

Consider a hemiellipsoidal emitter pin of height $h = 1000$nm and tip radius $R_a = 5$nm, mounted on a grounded conducting plane with anode kept at distance $D = 1100$nm from the grounded cathode plane. Thus $D = 1.1h$ or the anode is at a distance 100nm from the emitter apex.
At locations (i) $\rho_0 =0 $ and (ii) $\rho_0 = R_a/5$ on the surface of a hemiellipsoid, we compare Eq.~(\ref{eq:final}) with the results obtained using COMSOL in Figs.(\ref{fig:ellip_anode_0}) and (\ref{fig:ellip:anode_Ra_5}). The agreement is very good in the tunneling region, which implies that the form of external potential is unaltered even with the anode in the close proximity to the emitter. Thus, the external potential follows the same approximate universal trend observed when the anode is far away.

\subsection{Emitter array }

\begin{figure}[bh]
  \includegraphics[scale=.33]{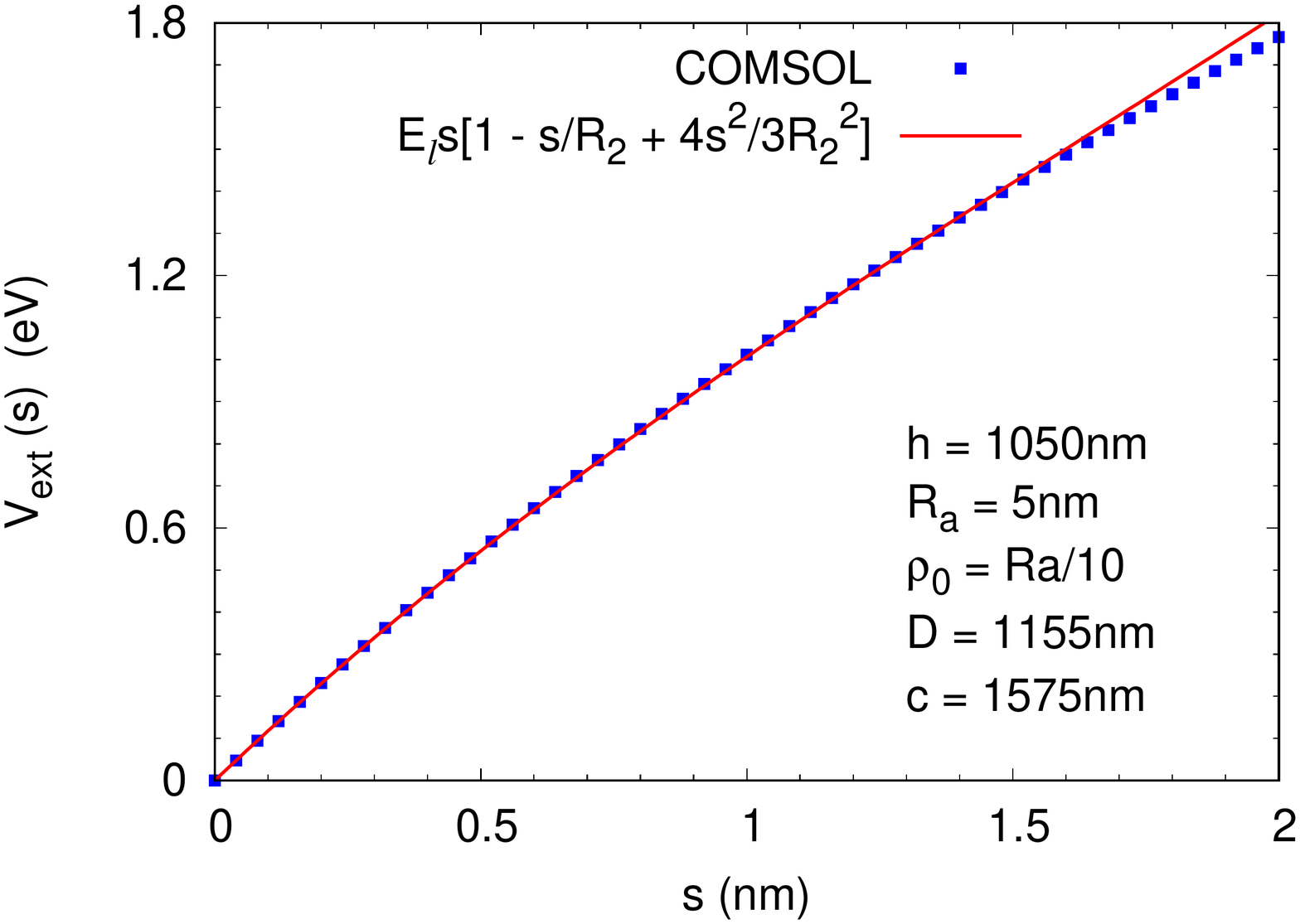}
  \vskip -0.25in
\caption{A comparison of the external potential variation at $\rho_0 = R_a/10$ for HECP emitters placed in a uniform square array of lattice constant  $c = 1.5h$, with the exact result generated using COMSOL (solid squares).}
\label{fig:HECP_array_Ra_10}
\end{figure}

\begin{figure}[th]
\includegraphics[scale=.33]{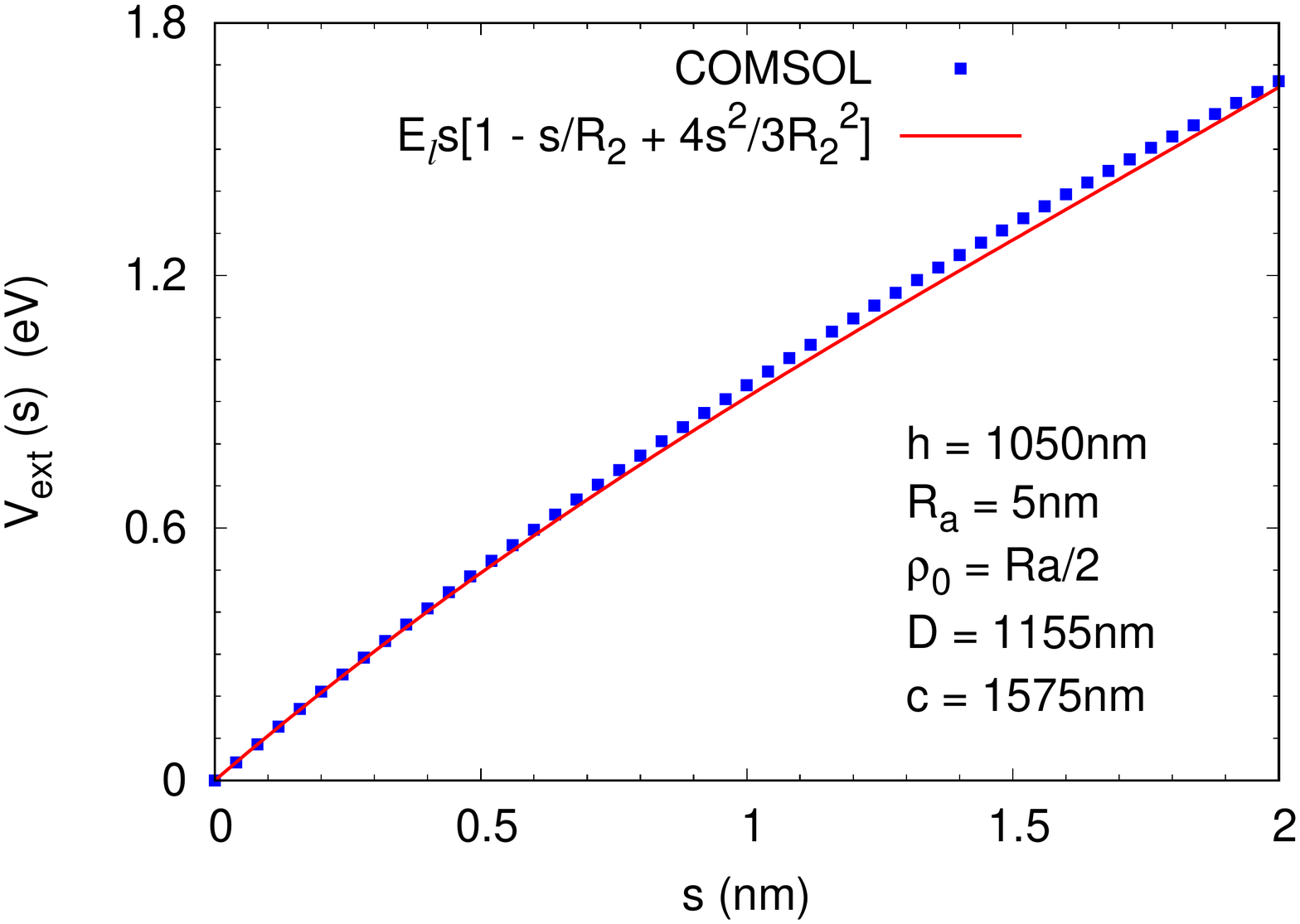}
\caption{ Same as that of Fig:(\ref{fig:HECP_array_Ra_10}) at $\rho_0 = R_a/2$. }
\vskip -0.25in
\label{fig:HECP_array_Ra_2}  
\end{figure}

So far, we have dealt with an isolated emitter with anode at close proximity as well as far away.
We now consider the case in which the emitter under consideration is a part of an infinite array. This situation arises when we are dealing with large area field emitters (LAFEs) used in high current applications. In emitter arrays, a notable effect is the reduction in the apex field enhancement factor ($\gamma_a $) compared to that of an isolated emitter. This is attributed to the shielding effects of the neighbouring emitters. The extent of shielding depends on the spacing and configuration of the other emitters in the array.

To study the nature of the external potential in an array, consider HECP emitters of total height $h = 1050$nm placed on a square lattice. The cylindrical post of height $1000$nm has a hemiellipsoid endcap of height $50$nm and tip radius $R_a = 5$nm. The lattice constant $c = 1.5h$. Thus, the square domain extends from
[-$c/2,c/2$] in both X and Y direction while the HECP emitter is placed at the origin. The anode is kept close  at a distance $D = 1155$nm. Thus, both shielding and anode-proximity effects are present in this case. The numerical result obtained using COMSOL is compared with Eq.~(\ref{eq:final}) at two locations (i) $\rho_0 = R_a/10$ (Fig~\ref{fig:HECP_array_Ra_10}) and (ii) $\rho_0 = R_a/2$ (Fig.~\ref{fig:HECP_array_Ra_2}). The agreement is found to be good in both cases.

Thus, irrespective  of the presence of anode or other emitters in close proximity, the form of external potential is approximately universal in the tunneling region.

\subsection{The Tunneling potential}

So far we have been discussing the effect of curvature on the potential due to an external applied field.
To put things in a proper perspective, it is instructive to look at the full tunneling potential, $V_T$

\be
\begin{split}
  V_T(s) = &~ \phi -eE_l s \left[ 1 - \frac{s}{R_2} + \frac{4}{3} \left(\frac{s}{R_2}\right)^2 \right] - \\
  &   \frac{e^2}{16\pi\epsilon_0 s(1+s/2R_2)} \label{eq:TP}
\end{split}
\ee

\noi
seen by the electrons emanating from the emitter tip. As in case of the external potential,
the image potential is also affected by the curvature of the tip. The last term in
Eq.~(\ref{eq:TP}) represents
the image-charge potential under a local spherical approximation\cite{db_image}.

\begin{figure}[h]
\vskip -0.6cm
\hspace*{-0.85cm}\includegraphics[scale=.365]{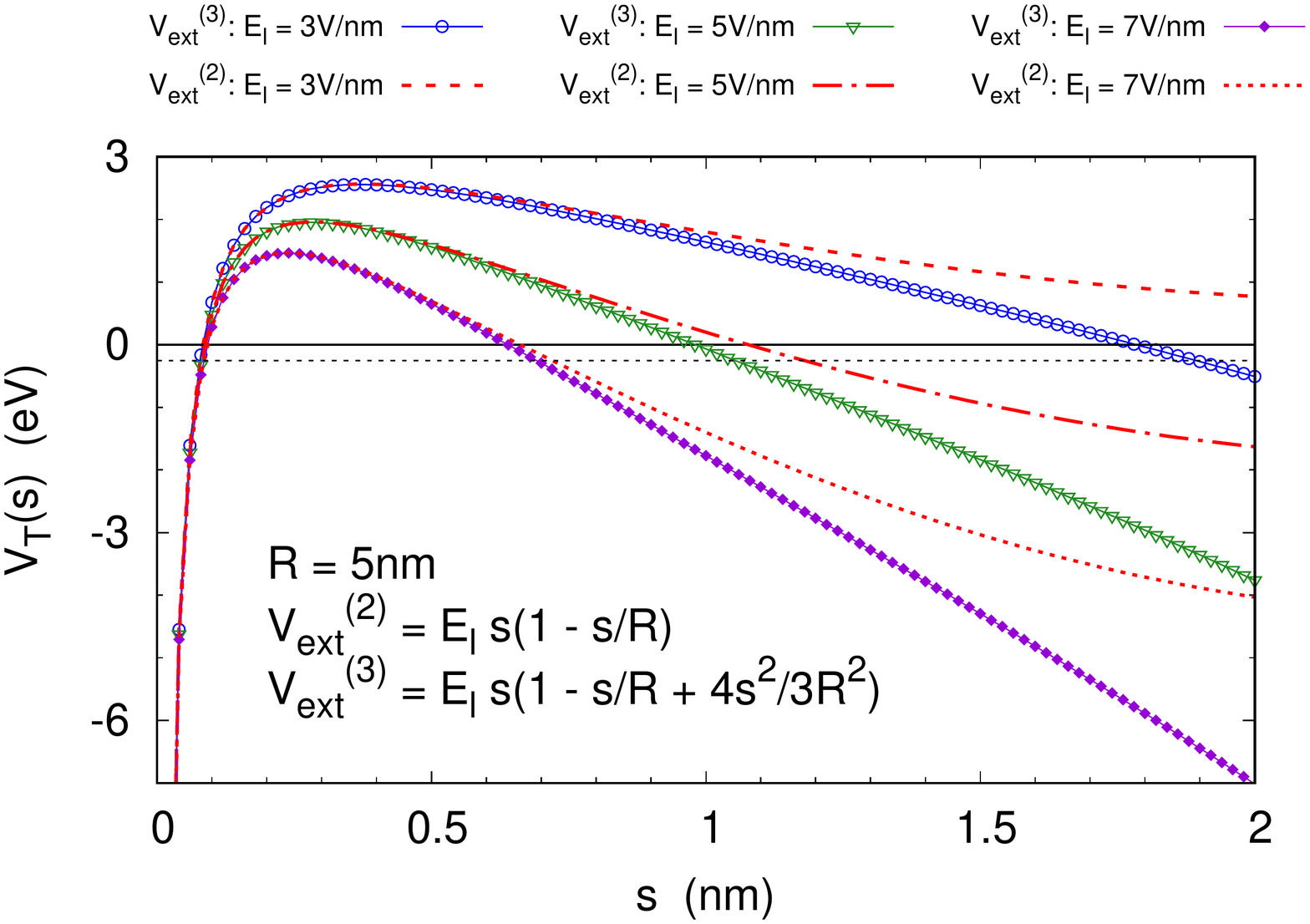}
\caption{The figure shows the comparison of tunneling potential for an HECP emitter at a point near the apex with $R = 5 $nm and for two possible external potential having (i) quadratic correction ($V_{ext}^{(2)}$) and (ii) cubic correction ($V_{ext}^{(3)}$) for three values of local field $E_l$. Here, the work function $\phi = 4.5$eV and $V_T = -0.25$eV is shown by dashed line (black).}
\label{fig:Tunnel_Pot}
\end{figure}

In order to determine the importance of the cubic order correction term in the external potential, we need to compare the tunneling potential with and without the cubic correction over typical tunneling distances involved at various values of the local field as shown in Fig.~\ref{fig:Tunnel_Pot}. The notations
$V_{ext}^{(2)}$ and $V_{ext}^{(3)}$ in Fig.~\ref{fig:Tunnel_Pot} represent the order of curvature correction
and are given by

\bea
V_{ext}^{(2)} & = & E_l s \left[1 - \frac{s}{R} \right] \\
V_{ext}^{(3)} & = & E_l s \left[1 - \frac{s}{R} + \frac{4}{3} \left(\frac{s}{R}\right)^2 \right]
\eea

\noi
with the local field at a point in the apex-neighbourhood denoted by $E_l$ and the local second principle radius of curvature
denoted by $R$. The value of $R$ is taken to be 5nm while three values of the local field $E_l$ are considered in Fig.~\ref{fig:Tunnel_Pot}.
In computing the tunneling potential, the work function is taken to be $\phi = 4.5$eV.
Note that at smaller fields, emission occurs close to the
Fermi level while at higher fields, the peak of the normal energy distribution shifts further away from the
Fermi level \cite{swanson,db_parabolic} (see for instance Fig.~4 in Ref.~[\onlinecite{db_parabolic}]).
Thus, while at $E_l = 7$V/nm the difference between $V_{ext}^{(2)}$ and $V_{ext}^{(3)}$ is small, at $E_l = 5$V/nm, the difference in tunneling
width is significant for $V_T = -0.25$eV.  Finally, at $E_l = 3$V/nm, removing the cubic term clearly makes the barrier unphysical.

Thus, the cubic order corrected external potential form given by Eq~(\ref{eq:final}) is a better approximation. Since a slight change in the tunneling potential can have a magnified impact on the value of transmission coefficient, our calculations show that it is profitable to use a cubic order corrected external potential at least when dealing with emitter tips having radius less than $20$nm at typical local fields of $3-10$ V/nm.

\section{Discussions and Conclusions}

We have shown, using the nonlinear line charge model, that the external potential
has an approximate universal form for sharp emitters irrespective of the particulars
of its shape or endcap geometry so long as it is smooth and parabolic in nature. The universal form given by Eq.~(\ref{eq:final}) holds not only on the axis of the emitter, but also away from the axis on the emitter surface. From the numerical studies we found that the Eq.~(\ref{eq:final}) can be applied to a wide generic class of simple as well as complex geometries which need not be well represented by the parabolic approximation for $\rho < R_a$. Also, Eq.~(\ref{eq:final}) was found to be applicable when the emitter is in close proximity to the anode or when it is part of an array.

By knowing the exact form of tunneling potential at all points on the surface of the emitter, one can  find the transmission coefficient using approximate techniques such as  WKB method, or by numerically solving\cite{db_vk} the Schr\"odinger equation or even using the transfer matrix method\cite{db_vk}. The total field-emission current from a sharp tip can thus be calculated from a detailed knowledge about tunneling potential. Fortunately, the form of the
tunneling potential appears to be approximately universal, independent of the emitter geometry. This should act as an incentive to find approximate analytical expressions for the field emission current density\cite{db_rr_2021} and even the net emitted current. 
      
\section{Acknowledgements}

The authors acknowledge useful discussions with Dr. Raghwendra Kumar.

\vskip 0.25 in
\noi
{\it Data Availability}: The computational data that supports the findings of this study are available within the article.

\section{References} 


\appendix*

\section{Bounds on correction terms}

In the following, we shall briefly estimate upper bounds for the correction terms
$\cc_0, \cc_1$ and  $\cc_3$ for  a typical emitter with $h/R_a = 1000$.

\subsection{The term $\cc_0$}

The correction term $\cc_0$ is

\be
\cc_0 = \int_0^L dt~\frac{f'(t)}{f(L)} \frac{t^3/(z^2 - t^2)^2}{L^3/(z^2 - L^2)^2}
\ee

\noi

Substituting $x = t/L$, and using Bernstein's inequality,

\bea
\cc_0 & =  & \int_0^1 dx~\frac{f'(x)}{f(L)} \frac{x^3(a^2 - 1)^2}{(a^2 - x^2)^2} \\
& \leq &  n \frac{\norm{f}}{f(L)} \int_0^1 dx~ \frac{x^3(a^2 - 1)^2}{(a^2 - x^2)^2 (1 - x^2)^{1/2}} \\
& \leq & n \left[ \frac{a^2-1}{2} + \frac{\pi}{4} \sqrt{a^2 - 1} \right] \\
& \approx & n \frac{\pi}{4} \sqrt{\frac{R_a}{h}}
\eea

\noi
where $a = z/L \approx h/L \approx h/\sqrt{h(h-R_a)} \approx 1 + R_a/2h$. Thus $\sqrt{a^2 - 1} \approx \sqrt{R_a/h}$.
Generally $\norm{f}/f(L) \leq 1$ since the line charge density peaks at $L$.
 For $n = 5$ and $R_a/h = 1/1000$, $\cc_0 \leq 0.124$.

\subsection{The term $\cc_1$}

The correction term $\cc_1$ is

\be
\cc_1 = \int_0^L dt~\frac{f'(t)}{f(L)} \frac{t/(z^2 - t^2)}{L/(z^2 - L^2)}.
\ee

\noi
which can in turn be expressed as

\bea
|\cc_1| & = & \int_0^1 dx~\frac{f'(x)}{f(L)} \frac{x(a^2 - 1)}{(a^2 - x^2)} \\
  & \leq & n \frac{\norm{f}}{f(L)} \int_0^1 dx~\frac{x(a^2 - 1)}{(a^2 - x^2)(1 - x^2)^{1/2}} \\
& \leq & n \frac{\pi}{2}  \sqrt{a^2 - 1} \\
& \approx & n \frac{\pi}{2} \sqrt{\frac{R_a}{h}}
\eea

\noi

For $n = 5$ and $R_a/h = 1/1000$, $\cc_1 \leq 0.248$.

\subsection{The term $\cc_3$}

The correction term $\cc_3$ is

\be
\cc_3 = \int_0^L dt~\frac{f'(t)}{f(L)} \frac{t^3/(z^2 - t^2)^3}{L^3/(z^2 - L^2)^3}
\ee
which in turn can be expressed as 

\bea
\cc_3 & =  & \int_0^1 dx~\frac{f'(x)}{f(L)} \frac{x^3(a^2 - 1)^3}{(a^2 - x^2)^3} \\
& \leq &  n \frac{\norm{f}}{f(L)} \int_0^1 dx~ \frac{x^3(a^2 - 1)^3}{(a^2 - x^2)^3 (1 - x^2)^{1/2}} \\
& \leq & n \frac{3}{8} \left[ (a^2 - 1) + \frac{\pi}{2} \sqrt{a^2 - 1}  \right] \\
& \approx & n \frac{3\pi}{16} \sqrt{\frac{R_a}{h}}
\eea

For $n = 5$ and $R_a/h = 1/1000$, $\cc_3 \leq 0.09$.

\end{document}